\documentclass[aps,prl,groupedaddress,twocolumn,showpacs]{revtex4}
\usepackage{graphicx}
\usepackage{latexsym}

\begin{document}

\title{Anomalous proximity effect in an inhomogeneous disordered superconductor}

\author{W. Escoffier, C. Chapelier, N. Hadacek and J-C. Vill\'egier}

\affiliation{CEA-DSM-DRFMC-SPSMS, CEA Grenoble, 17 rue des Martyrs, 38054 Grenoble Cedex 9, France.}

\date{\today}

\begin{abstract}

By combining very low temperature scanning tunneling microscopy and spectroscopy on a TiN film we have observed 
a non uniform state comprising of superconducting (S) and normal (N) areas. The local density of states displays a
spatial dependence between S and N different from the usual proximity effect. We conclude that mesoscopic fluctuations 
might play a major role in accordance with recent theories on superconductor-normal metal quantum transition. 
\end{abstract}

\maketitle

\begin{figure}
 \includegraphics[width=0.4\textwidth]{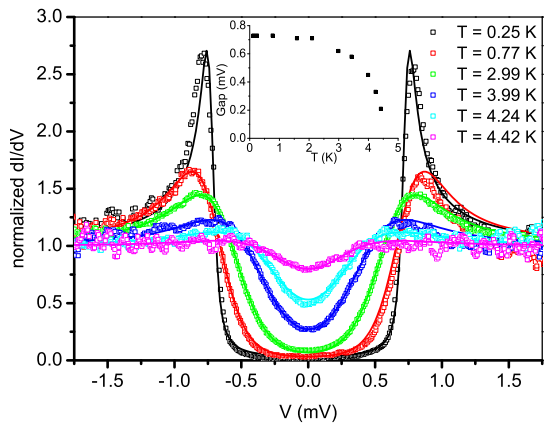}
 \caption{\label{gap-T}Several spectra taken at different temperatures at the same location and the corresponding BCS fits. 
Inset : temperature dependence of the BCS gap}
\end{figure}

In two dimensions, the superconductor-insulator transition (SIT) is traditionally described by two microscopic 
mechanisms \cite{Larkin}. In the first one , the Cooper pairs are all formed at T=0 and become localized through the SIT. 
This is often refered to as the \textquotedblleft bosonic\textquotedblright model and has been successfully to granular 
superconductors \cite{Goldman}. The onset of the superconductivity appears at a constant temperature but 
the transition becomes broader as the sheet resistance $R_\Box$ approaches $\frac{h}{4e^2}$. This broadening is 
associated with quantum fluctuations of the phase of the order parameter due to the competition between the charging 
energy of the superconducting grains and the Josephson coupling between them. 
The second mechanism attributes the weakening of the superconductivity to a disorder enhanced Coulomb repulsion. 
Pair breaking is considered simultaneously with the decreasing of both the superconducting critical temperature and 
the amplitude of the order parameter as the SIT is approached. This so-called \textquotedblleft fermionic\textquotedblright  
scenario usually describes homogeneous thin films which keep a sharp superconducting transition until
disorder becomes very close to a critical value \cite{Finkelstein}. By further increasing the disorder, by reducing the 
thickness or by application of a magnetic field, these films can be driven into an insulating state with localized 
electronic excitations. 
However, real samples may be neither purely bosonic or purely fermionic systems \cite{Sambandamurthy} 
and a unified theory describing these two limiting cases is still needed. Moreover, there are theoretical 
predictions \cite{SMT} and experimental observations \cite{Mason01,Ephron,Hsu,Chervenak} 
of an intermediate metallic phase right at the SIT. 
\\\indent 
In three dimensions, depending on the disorder, both a SIT 
and a superconductor-normal metal transition (SMT) are predicted at T=0 \cite{Larkin,Finkelstein} and have been 
experimentally observed in a field-tuned transition \cite{Gantmakher}. 
The existence of an intermediate metallic phase is often accompanied by a re-entrant 
temperature and field dependence of the electrical resistance $R$ of the film \cite{reentrance}. 
Actually, the competition between Cooper pairs and quasi-particule tunneling between superconducting grains or clusters 
can explain a nonmonotonic behavior of $R$. 
In these models, a zero resitance state appears when the network of superconducting islands percolate \cite{percolation}. 
Inhomogeneities in the morphology of granular films \cite{Gerber97} or strong statistical disorder in nominally 
homogeneous films can lead to the formation of these islands \cite{Fluct-stat}. 
Mesoscopic effects induced by local fluctuations in the density of randomly distributed impurities can also drive the film 
into a multiple reentrant field-tuned SMT \cite{Spivak95}. 
\\\indent 
A double reentrant behavior very close to a SIT has recently been observed in disordered TiN thin films \cite{Hadacek}. 
This SIT could be triggered either by a magnetic field or by increasing $R_\Box$. 
It is worth noticing that such a double reentrance has also been observed in Josephson junction arrays and highly 
granular superconductors \cite{reentrance2}. In this Letter we report spectroscopic measurements on a similar TiN film 
with a lower $R_\Box$ and in zero field, thus far away from the above mentioned SIT. 
Nevertheless, we show that disorder induced inhomogeneities already exist in such a film which cannot be detected 
by transport measurements.
\\\indent 
TiN was prepared by DC reactive magnetron sputtering at 350$^\circ$C on thermally oxidized Si substrates. By sputtering 
a Ti target at various nitrogen partial pressure in an argon-nitrogen gas mixture, different TiN$_\delta$ compounds with 
$0.7\leq\delta\leq1.2$ can be obtained \cite{Kang}. 
We have been able to vary the room temperature electrical resistivity between 80 $\mu\Omega$.cm and 1100 $\mu\Omega$.cm 
by changing the nitrogen flow rate from 40 to 200 sccm \cite{Nicolas}. In stoichiometric TiN films, 
the mean free path $l$ is mainly limited by the grain boundaries and one gets $l\simeq L_g$, 
where $L_g$ is the typical grain size. When the nitrogen flow rate is increased during the film preparation, 
Ti vacancies are introduced and one obtains overstoichiometric TiN$_\delta$ ($\delta>1$) with $l<L_g$. 
Moreover the Ti vacancies are not uniformly distributed inside the grains but are rather concentrated 
at their boundaries \cite{Patsalas}. 
The film studied here has an intermediate resisitivity of 270 $\mu\Omega$.cm, a mean free path $l=5\pm 1\;nm$ and a thickness of 100 nm \cite{Nicolas}. We estimate $L_g\simeq 20\; nm$ from STM images. 
It undergoes a sharp superconducting transition at $T_c$ = 4.68 K as detected by $R(T)$ experiments. 
\\\indent 
We have combined topography and spectroscopy measurements with a scanning tunneling microscope (STM) cooled down to very 
low temperature in a dilution refrigerator. In order to probe the local density of states (LDOS), a small AC modulation 
of 20 $\mu$V rms is added to the sample-tip DC bias voltage V  and the differential conductance $\frac{dI}{dV}$ is obtained 
with a lock-in amplifier technique. At any position above the sample V is ramped and the resulting $\frac{dI}{dV}$ (V) 
curve gives the LDOS around the Fermi energy. Several spectra are displayed in Fig.\ref{gap-T} at different temperatures . 
The superconducting gap values $\Delta (T)$ are obtained by convoluting a BCS density of states with a Fermi distribution 
function and are displayed in the inset. A weakly temperature dependent Dynes parameter in the range $\Gamma$ = 0.022mV to 0.027 mV 
is used to adjust the peak height \cite{Dynes}. The lowest indicated temperature of 250 mK is not the measured one 
which was 125 mK but rather the temperature needed to correctly fit the spectrum. This indicates that the energy 
resolution of our STM is probably limited by unfiltered electromagnetic radiations which heat the electrons. 
We find $\Delta$(T=0) = 0.73 mV and a ratio $\frac{\Delta}{kT_c}$ = 1.81 not far from the theoretical BCS value 
of 1.76. 

\begin{figure}
 \includegraphics[width=0.4\textwidth]{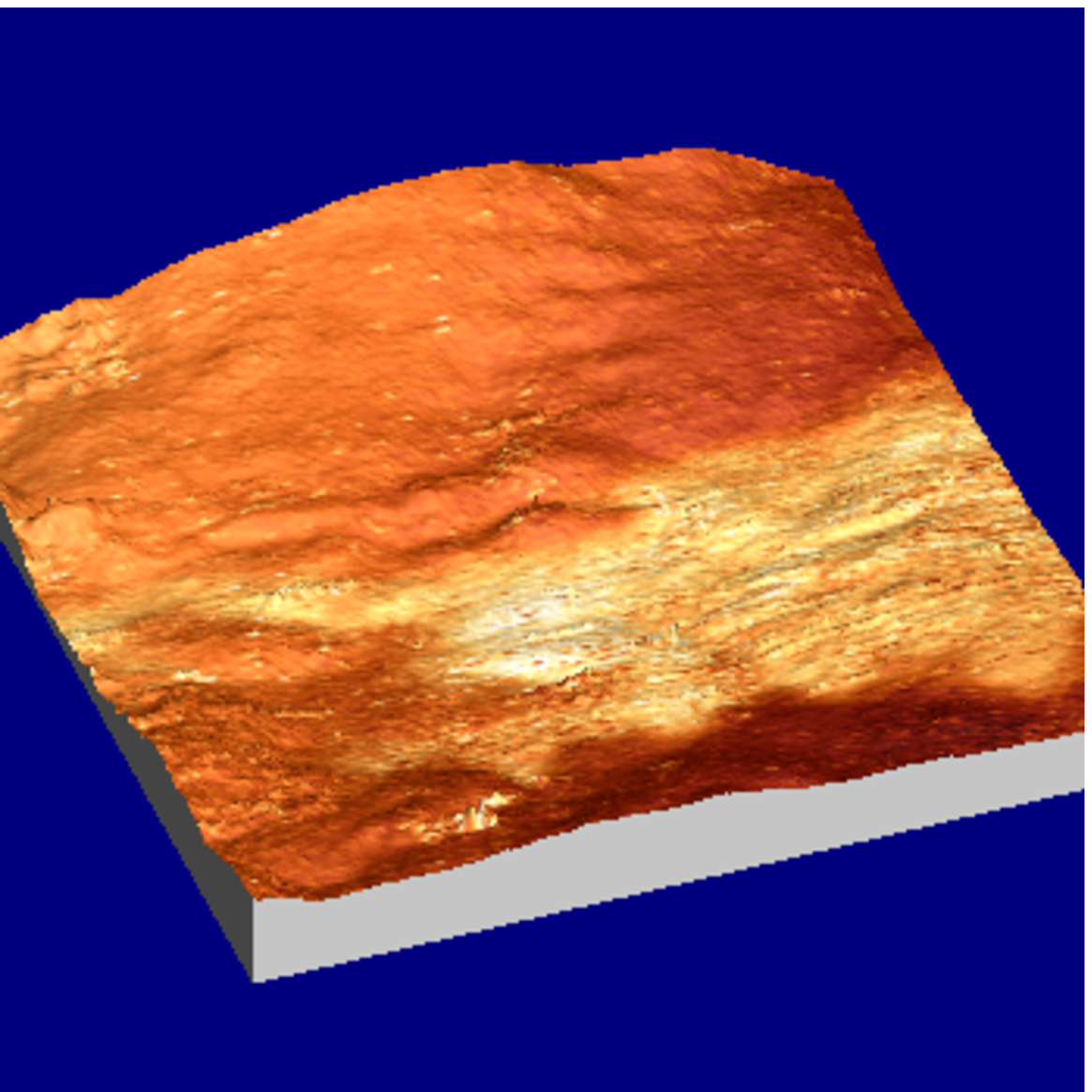}
 \includegraphics[width=0.4\textwidth]{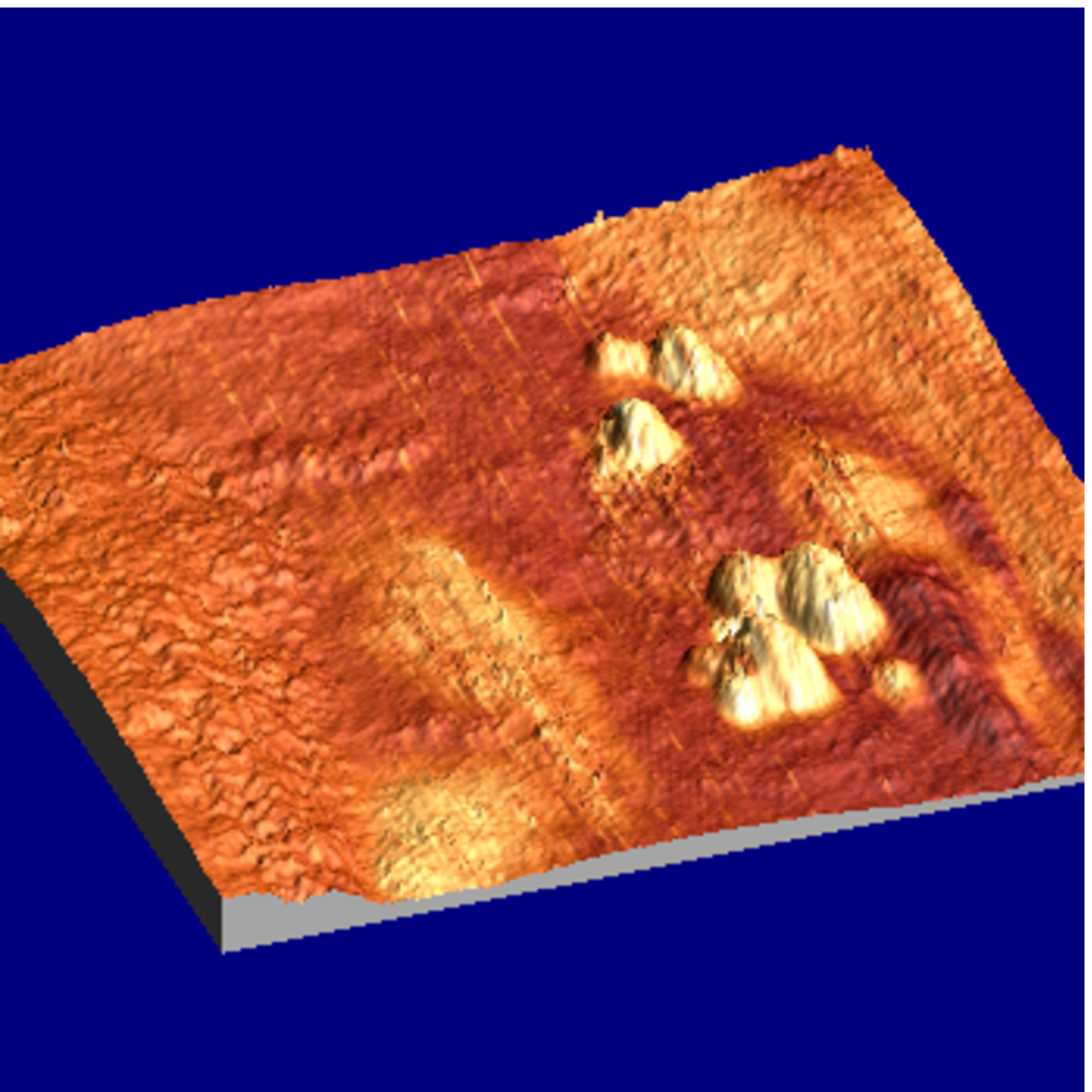}
 \caption{\label{photos}\rm\small 3D images of TiN : 400 x 400 nm$^2$ taken at 143 mK (top) and 250 x 250 nm$^2$ taken 
at 258 mK (bottom); the maximum height amplitude is 2nm for both pictures. The colouring maps the LDOS; the superconducting regions are yellow and the normal ones are orange. }
\end{figure}

However, these spectra are not observed everywhere on the surface of the TiN film and at other locations a normal-metal 
like flat LDOS is measured. In order to get an image of the superconducting and normal areas we set the bias 
to a voltage V$_0$ = 0.75 mV slightly above the BCS gap value. 
Two images were then recorded simultaneously : the topographic one Z(x,y) and the spectroscopic one $\frac{dI}{dV}$ (x,y) 
at a given energy eV$_0$. When the tip is scanning above a superconducting region, the differential conductance signal 
increases because of the peak in the LDOS at V$_0$. Inversely, a flat LDOS is characterized by a lower output 
of the lock-in amplifier. 
The two images can be merged into a 3D coloured picture as shown in Fig.\ref{photos} 
where the dark orange areas are normal while the light yellow ones are superconducting \cite{Nanotec}.

\begin{figure}
 \includegraphics[width=0.4\textwidth]{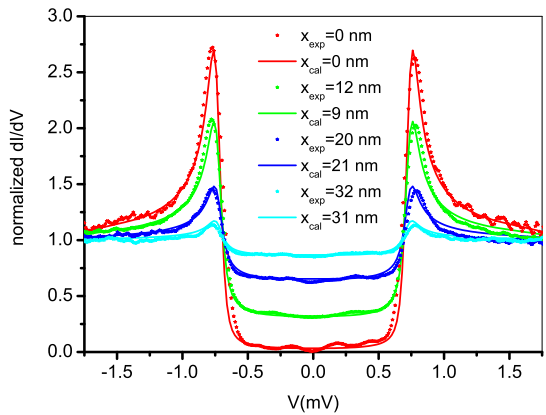}
 \caption{\label{interface}Spatial evolution of the LDOS between a superconducting and a normal region. 
The dotted lines are experimental data. The overall measured proximity length is $d_{SN} = 48 \;nm$. 
The solid lines are numerical fits with Eq.(\ref{lin}) which has no adjustable parameter and the boundary 
conditions Eq.(\ref{CI}).}
\end{figure}

Sometimes, isolated grains stand higher on top of the surface. They can be either normal or superconducting as is shown, respectively, in the upper and lower picture of Fig.\ref{photos}.
Once the electronic properties are identified everywhere on the scanned surface, it is then possible to measure 
the complete spectra along a line which crosses the boundary between a superconducting and a normal region. 
Several spectra along such a scan are displayed in Fig.\ref{interface}. 
Except for the trully gaped one which is labeled as the starting position $x_{exp} = 0$ on the S side, 
none of the spectra are BCS-like. The distance $d_{SN}$ over which the LDOS 
is varying depends on the local granularity. It can be as long as 50 nm when the surface is smooth. But if we scan across 
an isolated superconducting grain such as those in the bottom panel of Fig.\ref{photos}, the spatial variation of the LDOS is much more rapid and 
takes place within less than 10 nm, mainly inside the grain itself. However, these different spatial scales do not modify 
the form of the overall transition of the LDOS at the SN interface since the spectra evolve in the same manner between S 
and N wherever they are obtained on the surface of the sample.

The spatial dependence of the proximity effect at an SN interface can be described in the framework of the quasiclassical 
Green's functions by a complex pairing angle $\theta (E,{\bf r})$ \cite{Belzig}. The LDOS $n$ at a position $\bf r$ and 
for an energy $E$ from the Fermi level is related to $\theta$ by $n(E,{\bf r})=n_0Re[\cos\theta (E,{\bf r})] $. 
In the dirty limit, i.e. $l\ll\xi=\sqrt{\hbar D/2\Delta}$ ($D$ is the diffusion constant), 
$\theta(E,{\bf r})$ obeys the Usadel equation which in one dimension can be written as\cite{Usadel}:
\begin{eqnarray}
\frac{\hbar D}{2}\frac{\partial^2\theta}{\partial x^2} +[iE-\Gamma_{in}-2\Gamma_{sf}\cos\theta]\sin\theta & &\nonumber\\
+\Delta(x)\cos\theta &= &0, \label{green}
\end{eqnarray}
where $\Gamma_{sf}$ and  $\Gamma_{in}$ are the spin-flip and the inelastic scattering rates respectively. 
This description of the superconducting proximity effect at a mesoscopic scale has already been checked experimentally 
in SN heterojunctions with nanofabricated tunnel junctions and with STM experiments \cite{prox}. 
For infinite SN systems, it has been shown that the LDOS exhibits 
a V-shaped pseudo-gap with peaks below and above the Fermi level. These peaks are separated by an energy which 
decreases as a function of the distance from the interface in the normal side of the junction. 
Here, we observe instead U-shaped spectra with peaks that are pinned at the BCS energy $\Delta$ for any position between 
S and N. The LDOS is flat for energies smaller than $\Delta$ and increases as the tip is moved progressively away from the 
S region. We tried to fit these results using Eq.(\ref{green}) and a self-consistently 
determined space dependent order parameter $\Delta(x)$ \cite{DeGennes}.
However, no set of parameters was able to correctly reproduce even qualitatively the shape of the spectra. One of the 
possible reasons for this failure could be that our film does not fulfill the dirty limit condition. Actually, we have
$\xi =6.5\; nm$ not much bigger than $\l$\cite{Nicolas}. Moreover, since the scattering centers are concentrated at the grain boundaries, 
the electronic trajectories could be considered to be quasiballistic inside the grains. 
The projection in only one dimension of the Usadel equation could also be a too crude approximation.
\\\indent 
It is nevertheless striking than we can reproduce with a good accuracy our results if we assume the pairing angle varies linearly with distance : 
\begin{eqnarray}
\frac{\partial\theta}{\partial x} = \frac{\theta_{BCS}-\theta_N}{d_{SN}}\label{lin}
\end{eqnarray}
where $tan(\theta_{BCS}) = \frac{i\Delta}{E+i\Gamma}$ and $\theta_N = 0$. Eq.(\ref{lin}) has been integrated with 
the boundary conditions, 
\begin{eqnarray}
\theta(x=0)= \theta_{BCS} \quad and \quad \theta(x=d_{SN})=  0 \label{CI}
\end{eqnarray}
The numerical results are shown in Fig.\ref{interface}. We found an excellent agreement between the experimental data and 
the calculations without any adjustable parameters. $\Delta$ and $\Gamma$ are deduced from the spectrum taken above the BCS superconducting region and positions are chosen in order to match the measured LDOS at the Fermi energy. 
The slight differences between these positions and 
the measured ones reflect the granularity always present; the latter affects the measured value of the proximity length 
$d_{SN}$ and can also perturb the pure linear behavior of Eq.(\ref{lin}). Although, there is no obvious physical meaning behind this equation, we want to point out the universal character of this observed anomalous proximity effect since it does not 
depend on any physical properties of the material such as inelastic scattering times, conductivities or 
coherence lengths of S and N necessary to describe the usual proximity effect \cite{NbN}.
\\\indent We want to discuss now the physical origin of the superconducting and normal clusters. In STM experiments, surface 
contamination must be considered seriously as a possible artefact for samples exposed to air. Nevertheless, there are 
easily recognizable signs which indicate the presence of adsorbates : (i) they usually degrade the quality of the 
images. (ii) At moderately low temperature, when they are frozen out, their positions are revealed as bumps 
in the topographic images. (iii) At very low temperature, they generally exhibit the LDOS of an insulator and give 
dark areas in spectroscopic images. As seen in the pictures of Fig.\ref{photos}, no such effects are visible. 
The transition between superconducting and normal areas is in places very smooth with no change in the noise in the data. This proves that the inhomogeneities are intrinsic to the sample. However, we cannot rule out 
that because of oxidation of intergrain boudaries, for example, the surface of TiN itself is not representative of the bulk which could be more homogeneous. 
Another possible source of inhomogeneities, could come from spatial variations of the nitrogen concentration 
in TiN$_\delta$, since the BCS coupling constant, $\lambda$ depends on $\delta$ \cite{SupraTiN}. 
Following Ioffe and Larkin's pionnering work, strong statistical fluctuations of $\lambda({\bf r})$ 
can lead to an inhomogeneous system where superconductivity appears first in localized drops which percolate at $T_c$ 
even far away from the SIT threshold \cite{Fluct-stat}. 
A spatial distribution of the order parameter amplitude would give 
a broadened density of states as obtained by Hsu \textit{et al} with large area tunnel junctions \cite{Hsu}.
Here, we would  expect instead to observe well identified superconducting clusters with different BCS gap values.
It seems therefore more likely that our film is made of well coupled superconducting grains with a unique $\Delta$ embedded in a normal metal matrix where the Ti vacancies pile up.
According to Feigel'man \textit{et al} and Spivak \textit{et al}, quantum fluctuations in such a system 
make superconductivity unstable \cite{Feigelman-Spivak}. Spatial variations of the size and the concentration 
of grains as well as mesoscopic fluctuations of the intergrain conductance can drive the film locally into either 
a normal metal or a superconducting cluster. This is also consistent with the observed leveling off of the resistivity 
at very low temperature and with the reentrant field tuned transition observed in thinner TiN films. 
Indeed, when the film is closer to the percolation threshold of the superconducting network
its global resistivity is then governed by very few bottlenecks and the associated mesoscopic fluctuations. 
According to Spivak \textit{et al}, these mesoscopic fluctuations can give a multiple reentrant transition 
between superconducting and normal metallic states \cite{Spivak95}.
\\\indent
In summary, we have observed inhomogeneities in a disordered superconducting thin film of a nature similar to those invoked  to explain macroscopic properties \cite{Chervenak,Hsu,Mason01} and possibly described by recent fluctuations driven 
SMT models \cite{Feigelman-Spivak}. It is very likely that in order to understand the spatial dependence of the LDOS 
between superconducting and normal areas, one should go beyond the mean field Usadel equation : mesoscopic fluctuations could help explain the inhomogeneous superconducting state, the re-entrant field tuned transition in TiN and modify the proximity effect \cite{Fluct-meso}. 
\\\indent
We are grateful to M. Houzet for helping us in resolving self-consistently Usadel equation and to 
M. Sanquer who initiated these experiments.

\end{document}